%% file: Main.tex
\documentclass[journal=nalefd, manuscript=letter]{achemso}

\usepackage{chemformula} 
\usepackage[T1]{fontenc} 
\usepackage{setspace}
\usepackage{placeins}
\usepackage{booktabs}
\usepackage{hyperref}
\usepackage{caption}
\usepackage{subcaption}

\hypersetup{
  colorlinks   = true, 
  urlcolor     = blue, 
  linkcolor    = red, 
  citecolor   = blue 
}

\author{Hui Chen}

\author{Duncan T.L. Alexander}

\author{Cécile Hébert}
\affiliation[École Polytechnique Fédérale de Lausanne (EPFL)]
{Electron Spectrometry and Microscopy Laboratory (LSME), Institute of Physics (IPHYS), \'Ecole Polytechnique F\'ed\'erale de Lausanne (EPFL), 1015 Lausanne, Switzerland}
\email{cecile.hebert@epfl.ch}

\title[]{Leveraging Machine Learning for Advanced Nanoscale X-ray Analysis: Unmixing Multicomponent Signals and Enhancing Chemical Quantification}


\begin{document}

\begin{abstract}

Energy dispersive X-ray (EDX) spectroscopy in the transmission electron microscope is a key tool for nanomaterials analysis, providing a direct link between spatial and chemical information. However, using it for precisely determining chemical compositions presents challenges of noisy data from low X-ray yields and mixed signals from phases that overlap along the electron beam trajectory. Here, we introduce a novel method, non-negative matrix factorisation based pan-sharpening (PSNMF), to address these limitations. Leveraging the Poisson nature of EDX spectral noise and binning operations, PSNMF retrieves high quality phase spectral and spatial signatures via consecutive factorisations. After validating PSNMF with synthetic datasets of different noise levels, we illustrate its effectiveness on two distinct experimental cases: a nano-mineralogical lamella, and supported catalytic nanoparticles. Not only does PSNMF obtain accurate phase signatures, datasets reconstructed from the outputs have demonstrably lower noise and better fidelity than from the benchmark denoising method of principle component analysis.

\end{abstract}

\section{MAIN TEXT}
An energy-dispersive X-ray spectrum (EDX spectrum) is obtained by bombarding a sample with a beam of high energy electrons which induces the emission of characteristic X-rays from the sample's atoms~\cite{williams2009x}. These emitted X-rays are then measured and analyzed, both to identify the elements present in the material and to quantify their relative abundance. By combining EDX with scanning transmission electron microscopy (STEM-EDX), a sub-nm or even atomic spatial resolution image can be obtained for element mapping~\cite{Allen2012}. As a result, this technique is widely employed for the chemical characterization of nanomaterials, playing an essential role in understanding their properties and performance by determining the elemental compositions of their constituent phases or compounds~\cite{li2017surface,divitini2016situ,el2021atomic,shim2021utilization,krans2020stability}. Nevertheless, the technique poses challenges. When quantifying nanomaterials with STEM-EDX, the signal-to-noise ratio (SNR) of the spectral signal is often limited due to two main factors: first, the very small sample volume interacting with the focused electron beam; second, limitations on the electron dose to avoid beam-induced damage. Moreover, it is difficult to quantify heterogeneous nanostructures or nanocomposites using STEM-EDX when different phases overlap along the trajectory of the electron beam through the sample, resulting in EDX signals that derive from a mixture of phases. For precise compositional determination of each phase, it is vital to separate these mixed signals.

To address these challenges, machine learning algorithms are increasingly used in the analysis of STEM-EDX mapping data. Among these, principal component analysis (PCA)~\cite{jolliffe2002principal} has arguably become the \textit{de facto} ``gold standard'' for denoising datasets~\cite{moreira2022improving,potapov2019optimal,potapov2017enhancement}. Nevertheless, it is prone to introducing artifacts into reconstructed data resulting from its inability to ensure signal positivity~\cite{potapov2017loss,barbier2023fundamental}. Independent component analysis (ICA)~\cite{hyvarinen2000independent} and non-negative matrix factorization (NMF)~\cite{lee2000algorithms} are in turn commonly utilized for identifying components from mixed or overlapped phases~\cite{de2011mapping,rossouw2016blind,jany2017retrieving,jia2022machine}. However, these methods also have limitations: data noise can lead to incorrect outputs~\cite{wei2020overview,muto2020application}; and, if the chemical signals from individual phases are too similar, the capacity of ICA and NMF to disentangle them can prove poor~\cite{somers2011endmember,miao2007endmember,wang2013endmember}.

Pan-sharpening (PS) is a data-fusion technique widely used in satellite imaging~\cite{javan2021review}, which combines the high spatial resolution of a panchromatic image with the color information from a multispectral image to produce a single, high-resolution, color-enhanced image. Recently, Borodinov et al. adapted this concept for the enhanced analysis of STEM electron energy-loss spectroscopy (EELS) plasmonic datasets~\cite{Borodinov2021}. Similarly to the original PS methodology, their implementation depends on acquiring two datasets, one having high spectral fidelity but few spatial pixels, and the other having high spatial resolution but poor spectral signal. Now considering STEM-EDX, one specificity of the datasets is that the spectral noise has a Poissonian nature~\cite{bell2003energy,hasinoff2014photon}. A dataset having high spatial resolution but poor spectral signal can therefore be transformed into one having high spectral fidelity but fewer spatial pixels by simple spatial binning, all while keeping a Poissonian noise structure. Our new method, which we call PSNMF (NMF-based pan-sharpening), combines such binning operations with PS-inspired data fusion to leverage the strengths of NMF, notably its non-negativity and component separation capabilities. With this, we achieve both a highly effective unmixing of signals from overlapped phases and a very efficient signal denoising. 

The exact workflow is presented in Figure~\ref{fig:PSNMF}. We refer to the original dataset as HR-LS (high-resolution, low-signal). In this dataset, the EDX spectrum of each single pixel has a very low SNR given that, typically, it has just one to a few hundred X-ray counts when summed over its whole spectral energy range. From this HR-LS dataset, we perform $b\times b$ pixel spatial binning to generate a new dataset. This binning, while reducing spatial resolution, improves the SNR in each pixel's spectrum. It does so by increasing the signal in each spectrum by a factor of $b^2$. Since Poisson noise is proportional to the square root of the counts, the SNR per pixel is therefore improved by a factor of $b$. The binned dataset is referred to as LR-HS (low-resolution, high-signal). By strategically fusing the algorithmic decomposition outputs from the complementary datasets, HR-LS and LR-HS, we will reconstruct a high-quality dataset, denoted as HR-HS (high-resolution, high-signal). While PS fusion can be implemented through several methods, here we combine it with NMF, due to its ability to unmix signals while ensuring a physically-correct non-negativity that helps produce meaningful outputs. In all, the PSNMF procedure comprises four steps. These steps are now described, while a mathematical description of them is given in Supporting Information, Note 1.

\begin{enumerate}
\item A spatial binning of the original HR-LS dataset by a factor $b$ is applied to create a LR-HS dataset. $b$ can be large, since it does not affect the spatial resolution of the final reconstructed dataset. As a rule of thumb, the maximum $b$ is that which preserves individual spatial phase structures as distinctly as possible, without introducing additional mixing through binning. It is advisable to test several values of $b$, up to this maximum.

\item A first NMF decomposition is applied to the binned LR-HS dataset, producing spectral components with significantly improved accuracy when compared to an NMF decomposition performed on the original, noisier HR-LS dataset.

\item A second NMF decomposition is executed on the original HR-LS dataset to recover high spatial resolution abundance maps for the components. Critically, the spectral components derived from the first NMF decomposition are used to initialize the algorithm. Given that NMF operates as a heuristic algorithm and is, for noisy datasets, highly sensitive to initial values, this greatly improves the decomposition result when compared to that from a random initialization on the same dataset.

\item Finally, we fuse the high-accuracy spectral components delivered by the first NMF decomposition with the high spatial resolution abundance maps obtained from the second NMF decomposition. This produces a high quality reconstructed dataset, having a greatly enhanced SNR for the EDX signal while retaining the full spatial resolution of the original dataset.

\end{enumerate}

\input{Fig/Fig1}

We first illustrate and quantitatively test the capabilities of PSNMF by applying it to synthetic datasets having a known ground truth. Our choice of test case for constructing the synthetic dataset was based on the following considerations. First, we wanted a case that represents usual challenges encountered in the chemical analysis of samples obtained by solidification processes, as are typical for many areas of geology and materials science. Second, we selected a sample type that we have extensively studied, such that, with high confidence, we can produce a very realistic synthetic dataset. We therefore simulate a STEM-EDX dataset that mimics those acquired from mineral assemblages that are solidified under high pressure and temperature conditions for the study of Earth's deep mantle~\cite{nabiei2021investigating}. It comprises three distinct mineral phases that form and segregate with nanoscale spatial distributions: bridgmanite, ferropericlase, and calcium perovskite~\cite{hirose2017perovskite}. Bridgmanite (Brg), a silicate perovskite often denoted as \ch{(Mg,Fe)SiO3}, stands as the predominant mantle mineral. Notably, it may incorporate minor quantities of Al and Ca and also trace quantities of rare-earth elements such as Nd and Sm~\cite{liebske2005compositional}. Ferropericlase (Fp), chemically expressed as \ch{(Mg,Fe)O}, constitutes the second most abundant mineral in the lower mantle. The mineral with the smallest volume fraction is calcium perovskite (CaPv). It has a chemical formula of \ch{CaSiO3}, and acts as the primary host for rare earth elements within the mantle. The analysis of these mineral phases greatly enhances our understanding of Earth's geochemical evolution~\cite{corgne2005silicate,walter2004experimental}. 

Table~\ref{tab:comp1} presents the exact compositions set for Brg, Fp and CaPv for creating the synthetic dataset. Figure~\ref{fig:C147} shows their designed spatial distributions (``Ground truth'' column), which are based on those previously identified from the in-depth investigation of an experimental sample~\cite{chen2024unmixing}. As seen in the figure, there are many regions in which different phases overlap spatially. Together with the compositional overlaps seen in Table~\ref{tab:comp1}, this creates a basis that is typically challenging for effective data decomposition into correct phase signatures. From the phase compositions and distributions, a noise-free STEM-EDX dataset of $180\times 180$ pixels size is simulated using the ``espm'' (Electron Spectro-Microscopy) open-source Python library~\cite{Teurtrie2023}. Using the same software package, Poisson sampling is then applied to generate two datasets with realistic spectral noise. One dataset exhibits a moderate SNR -- corresponding to what can typically be achieved experimentally on such samples – while the other has an extremely low SNR. In the medium SNR dataset, the average X-ray count per pixel is 147. In contrast, the extremely low SNR dataset has an average of 18 X-ray counts per pixel. The raw elemental maps derived from these two synthetic datasets are provided in the Supporting Information Figure S1 and Figure S2. Before applying PSNMF to the synthetic datasets, we conducted PCA to determine the appropriate number of components for the NMF decomposition. In agreement with the three phases used to generate the datasets, the scree plots, included in the Supporting Information Figure S3, indicate the presence of three components, albeit with varying variances. 

\begin{table}[htbp]
    \small
    \centering
    \begin{tabular}{llllllllllll}
    \hline
        & Atomic \% & Mg & Si & Al & Ca & Fe & Nd & Sm & U  & O  & Cu \\ \hline
        & Brg & 18.50 & 18.20 & 1.50 & 1.00 & 1.00 & 0.10 & 0.10 & - & 60.00 & - \\ 
        Phase & Fp & 39.00 & - & - & - & 10.00 & - & - & - & 50.00 & 1.00 \\ 
        & CaPv & - & 17.50 & 2.50 & 16.20 & 1.00 & 1.00 & 1.00 & 1.00 & 60.00 & - \\ \hline       
        & PSNMF1 & 18.49 & 18.10 & 1.47 & 1.01 & 0.99 & 0.10 & 0.10 & - & 59.72 & - \\ 
        Medium SNR & PSNMF2 & 39.53 & - & - & - & 10.46 & - & - & - & 48.84 & 1.16 \\ 
        & PSNMF3 & - & 17.42 & 2.20 & 16.57 & 0.98 & 0.99 & 0.96 & 0.99 & 59.81 & - \\ \hline
        & PSNMF1 & 18.47 & 18.29 & 1.43 & 1.01 & 0.99 & 0.10 & 0.12 & - & 59.55 & - \\ 
        Low SNR & PSNMF2 & 39.65 & - & - & - & 10.36 & - & - & - & 48.63 & 1.26 \\ 
        & PSNMF3 & - & 17.43 & 2.41 & 15.47 & 1.05 & 0.99 & 0.91 & 1.08 & 60.25 & - \\ \hline
    \end{tabular}
\caption{The compositions of the three simulated phases utilized in generating the synthetic datasets and the compositions of PSNMF components when applied to the two synthetic datasets, using $b=12$ for the medium SNR dataset and $b=15$ for the low SNR dataset.}
\label{tab:comp1}
\end{table}

First, we employ PSNMF on the dataset with medium SNR, with incremental binning factors ($b = 2, 4, 12, 15, 30$). Upon reaching $b = 12$, PSNMF demonstrates an effective phase unmixing. This can be seen visually in Figure~\ref{fig:C147}, where the resulting spectral components (PSNMF\#) are compared to the spectra of the ground truth phases. As comparison, we also plot the results of a standard NMF decomposition performed on the same dataset. Qualitatively, the PSNMF spectral match is substantially improved compared to that of NMF, particularly for the phase with the lowest volume fraction (CaPv) where the NMF3 spectrum shows a large deviation from the ground truth. To assess this quantitatively, we use a spectral angle metric~\cite{singh2021review}, defined as:
\begin{equation}
    \label{eqn:angle}
    \alpha = \arccos{\left(\frac{\boldsymbol{v}_1 \cdot \boldsymbol{v}_2}{||\boldsymbol{v}_1|| \times ||\boldsymbol{v}_2||}\right)}
\end{equation}

where $\boldsymbol{v}_1$ and $\boldsymbol{v}_2$ are two spectral vectors of the same dimension. A smaller angle indicates a higher degree of spectral similarity; an angle of \(0^\circ\) indicates an exact match, while an angle of \(90^\circ\) implies no similarity. The close match of PSNMF1, PSNMF2 and PSNMF3 to the ground truths is confirmed by their respective spectral angles of \(0.25^\circ\), \(1.52^\circ\) and \(1.73^\circ\). In comparison, as seen in Figure~\ref{fig:C147}, the spectral angles for standard NMF are far greater, with values of \(5.81^\circ\), \(4.30^\circ\) and \(64.15^\circ\), respectively. The last value in particular confirms that standard NMF fails to correctly extract the minor phase CaPv. It is also evident that NMF overestimates Ca in Brg and underestimates O in Fp, as shown in the insets of the spectra in Figure~\ref{fig:C147}. This discrepancy does not occur for the components retrieved with PSNMF. Additionally, we quantify the three PSNMF components and list their compositions in Table~\ref{tab:comp1}. Comparison to the ground truth compositions shows that, impressively, even trace elements within Brg, such as Nd and Sm, are accurately quantified. These elemental quantifications were performed using the Cliff-Lorimer ratio method~\cite{cliff1975quantitative}, employing k-factors derived from X-ray emission cross-sections that were generated through state-of-the-art calculations using the ``emtables'' (Electron Microscopy Tables) library~\cite{Teurtrie2023}. One should note that the k-factors are the same as those used for constructing the synthetic datasets, hence discrepancies between quantification after reconstruction of a noisy dataset can only be explained by a failed matrix decomposition.

\input{Fig/Fig2}
\FloatBarrier

Under the PSNMF methodology, the second NMF decomposition yields the high spatial resolution components, as shown in the PSNMF maps in Figure~\ref{fig:C147}. Similarly to the spectra, these are compared to the abundance maps from a standard NMF decomposition. It is immediately seen that PSNMF retrieves phase distributions with much higher fidelity and lower noise than the standard NMF. This is particularly true for CaPv, where standard NMF largely fails to retrieve the phase distribution. To compare outputs quantitatively, we use a mean squared error (MSE) metric~\cite{singh2021review} that measures the accuracy of the retrieved components compared to ground truth phase distribution maps. MSE is defined as:
\begin{equation}
    \label{eqn:mse}
    \text{MSE} =  \frac{||\boldsymbol{M}_1 - \boldsymbol{M}_2||^2}{P_x P_y}
\end{equation}

where $\boldsymbol{M}_1$ and $\boldsymbol{M}_2$ are two maps of dimension $P_x \times P_y$ respectively representing the retrieved and ground truth abundance maps of a phase. The MSE can take values between 0 and 1, where 0 represents a perfect agreement between the two maps. The MSE values reported in Figure~\ref{fig:C147} confirm the remarkable ability of PSNMF to correctly retrieve the phase distributions. This is not the case for standard NMF, whose MSE values are an order of magnitude greater for the Brg and CaPv phases.

Having studied the power of PSNMF for phase unmixing, we now consider its use for dataset denoising. The spectral and spatial components generated by PSNMF are used to reconstruct a high-quality STEM-EDX dataset, from which elemental maps are extracted. As comparison, we apply PCA to the original dataset, since this remains the ``gold standard'' tool for hyperspectral data denoising. The dataset is then reconstructed using the first three components of the PCA decomposition. From the raw and reconstructed datasets, Figure~\ref{fig:C147} compares the resulting maps for a trace element, Nd. In the raw Nd map, because of the low spectral signal, noise is prominent, and the true spatial distribution cannot be distinguished. While PCA enhances the Nd map by decreasing noise and revealing the basic Nd distribution, the PSNMF-denoised Nd map clearly exhibits the least noise and closest fidelity to the ground truth map. The comparison of all elemental maps in the Supporting Information Figures S4 and S5 confirms the superior denoising by PSNMF, across the sample's full elemental range.

We now extend our evaluation of the PSNMF methodology by testing it on the second synthetic dataset having a very low SNR. The results are summarized and presented in Figure~\ref{fig:C15}. As before, we compare outputs to those of standard NMF and PCA. From visual inspection of the results, and the spectral angles of \(29.69^\circ\) to \(75.41^\circ\), it is evident that, because of the high noise level, NMF fails to correctly retrieve any of the phases. However, by applying $b = 15$, PSNMF yields phase unmixing results that are highly satisfactory given the poor quality of the initial dataset, with spectral angles ranging from just \(0.64^\circ\) up to \(5.04^\circ\). Remarkably, as seen in Table~\ref{tab:comp1}, quantification of the three PSNMF component spectra yields compositions that show minimal deviation from the designed composition. Therefore, we find that PSNMF is extremely effective for unmixing spatially overlapping and chemically similar phases even in the presence of high noise levels. As before, we use the three PSNMF components to reconstruct the STEM-EDX dataset. From this, Figure~\ref{fig:C15} shows the extracted Nd elemental map, compared to ground truth, raw and PCA-denoised maps. While the raw Nd map exhibits only noise, the map from the PSNMF-reconstructed dataset once again retrieves the true Nd distribution, with superior denoising compared to the map generated by PCA. A full comparison of ground truth, raw, and PCA- and PSNMF-denoised elemental maps can be found in Figures S6 and S7 of the Supporting Information.

\input{Fig/Fig3}
\FloatBarrier

Having quantitatively tested PSNMF on simulated datasets, we now validate its effectiveness on experimental data. The first sample studied is equivalent to that represented by the simulated datasets, having a nanometric distribution of Fp and CaPv phases in a Brg matrix. The sample was prepared using focused ion beam milling from a mineral assemblage synthesized under the conditions found in literature~\cite{nabiei2021investigating}. The acquired dataset has a spatial size of $512\times 512$ pixels and an average of 120 X-ray counts per pixel, therefore closely resembling the counting statistics of the medium SNR synthetic dataset. Figure~\ref{fig:C120} presents the results of PSNMF on this dataset, again compared to standard NMF. Note that, while we cannot compare the component outputs to any ground truth maps, we are able to compare to ground truth spectra, which we have calculated from the original dataset using a combination of NMF-derived masking and compositional prior knowledge, as described in our previous work~\cite{chen2024unmixing}. We conducted PSNMF experiments with binning factors $b = 2, 4, 8, 16, 32$; out of these, $b = 8$ gives the lowest average spectral angles, which are respectively just \(0.15^\circ\), \(0.96^\circ\), and \(2.64^\circ\) for Brg, Fp and CaPv. Given that, unlike the method used to obtain the ground truth spectra~\cite{chen2024unmixing}, these high fidelity spectra are obtained from spatially- and chemically-overlapped phases without recourse to any prior knowledge about phase-distinguishing elements, this demonstrates the powerful capability of our approach. Figure~\ref{fig:C120} further demonstrates how our methodology identifies clear distribution maps of all three phases. In contrast, while standard NMF retrieves Brg and Fp distributions, with moderate spectral angles of \(5.09^\circ\) and \(8.94^\circ\), it fails to retrieve the CaPv component, both spatially and spectrally (spectral angle = \(71.29^\circ\)). Turning to denoising, Figure~\ref{fig:C120} shows that the map of Nd from the PSNMF-reconstructed dataset is very clear, even though it is a trace element in the Brg matrix (at a level of approximately 180 ppm). Further to this, we present spectra integrated from the same Brg-located $4\times 4$ pixels in the raw, PCA-reconstructed and PSNMF-reconstructed datasets, to compare the $L\alpha$ peaks of trace rare earth elements Nd and Sm. The raw spectrum does not show any spectral shape, nor evidence of Nd and Sm. The PCA spectrum shows some spectral shape, and indicates some presence of Sm. However, clear Nd and Sm peaks are revealed using PSNMF. Moreover, this spectral signature matches well a Brg ground truth spectrum, that was itself determined by integrating $1.6\times10^5$ pixel spectra, thereby vindicating the improved denoising of the PSNMF over the PCA processing.

\input{Fig/Fig4}
\FloatBarrier

The mineralogical sample represents a major category of samples analysed by STEM-EDX: lamellae of uniform thickness that are prepared from bulk materials using ion beam milling and/or mechanical thinning. We now test the effectiveness of PSNMF on the other major category of analysed samples: nanostructures of irregular thickness that are supported on an amorphous carbon film. As a test object, we take a sample of Cu$_2$O nanocubes of $\sim$20 nm size decorated with $\sim$3--4 nm diameter Au nanoparticles, that is being developed as an efficient catalyst for electrochemical CO$_2$ reduction~\cite{Rettenmaier2023}. Like many heterogeneous catalysts, characterising and quantifying the size and distribution of the nanoparticles is of high scientific utility. However, as shown in Figure~\ref{fig:NPs}, while the larger Au nanoparticles are visible in the raw Au EDX map, their edges are indistinct, as for instance compared to the high angle annular dark-field STEM image shown in Supporting information Figure S8. Further, many of the smallest Au nanoparticles cannot be distinguished at all. Both of these problems derive from the low SNR inherent for such small particles, that have very limited volumes for EDX signal generation.

Now applying PSNMF to the EDX dataset with $b = 16$, Figure~\ref{fig:NPs} shows that our method correctly separates the catalytic structures into their separate phases, producing low-noise abundance maps and component EDX spectra that correspond to the Cu$_2$O nanocubes and the Au nanoparticles. From the Au abundance map alone, it is possible to quantify the sizes of the Au nanoparticles. As shown in Figure S9 in Supporting Information, two other phases are also identified, in turn corresponding to the amorphous carbon support film and a silica contamination. By reconstructing the EDX dataset using PSNMF, the right-hand column of Figure~\ref{fig:NPs} illustrates that we now achieve an elemental Au EDX map of low noise and high quality. Indeed, even Au nanoparticles of only 1.8 nm diameter are distinct from the background. This test case demonstrates that PSNMF is also an effective tool for the denoising and quantitative analysis of STEM-EDX data acquired from heterogeneous, supported nanostructures. 

\input{Fig/Fig5}

\FloatBarrier

In conclusion, modern analytical transmission electron microscopes equipped with fast, single or multi-segment silicon drift EDX detectors have revolutionized nanoanalytics by giving researchers the capability to generate high pixel density elemental maps in a short acquisition time. However, quantitative interpretation of these results is commonly challenging, owing to a high level of spectral noise, together with EDX phase signals that are mixed together. We have presented PSNMF, a novel machine learning methodology for addressing both of these challenges, at the same time. By applying PSNMF to realistic synthetic datasets, we have proven its superior performance compared to benchmark algorithms. We have further illustrated its effectiveness on two experimental use-cases: first, a thinned lamella of a nanoscale multi-phase system; second, heterogeneous nanoparticles supported on an amorphous carbon film. The high quality of phase decomposition and high fidelity of the denoised maps prove the strong potential of PSNMF for improving STEM-EDX analytics across a variety of nanoscience domains.

\begin{acknowledgement}
We acknowledge the Interdisciplinary Centre for Electron Microscopy (CIME) at EPFL for providing access to their electron microscopy facilities. Clara Rettenmaier, See Wee Chee and Beatriz Roldan Cuenya of the Fritz-Haber-Institute of the Max-Planck Society, Berlin, are thanked for providing the Au--Cu$_2$O catalyst sample; we acknowledge the Max Planck-EPFL Center for Molecular Nanoscience and Technology for faciliting this collaboration. Lastly, we thank James Badro and Stephan Borensztajn of LabEx UnivEarthS at the Institut de Physique du Globe de Paris (IPGP), Paris, for synthesising and preparing the mineralogical sample. 

\end{acknowledgement}

\clearpage
\bibliography{psnmf_bib}

\clearpage

\setcounter{figure}{0}
\renewcommand{\figurename}{Fig.}
\renewcommand{\thefigure}{S\arabic{figure}}

\section{Supporting Information for ``Leveraging Machine Learning for Advanced Nanoscale X-ray Analysis: Unmixing Multicomponent Signals and Enhancing Chemical Quantification''}

\section*{Note 1. Mathematical description of PSNMF}
We begin with the original noisy HR-LS (high-resolution, low-signal) dataset ($Y$) of size ($y$, $e$), where $e$ is the number of energy channels (spectral features) and $y$ is the number of pixels (spatial dimensions). Through spatial binning, we create the LR-HS (low-resolution, high-signal) dataset ($X$) of size ($x$, $e$), where $x$ is the number of pixels and is calculated as $x = \frac{y}{b^2}$, with $b$ being the bin size (e.g., a bin size of 4 results in $x = \frac{y}{16}$). We aim to combine the beneficial properties of $Y$ and $X$ into a new dataset $Z$. $Z$ is a matrix of size ($y$,$e$) that possesses improved SNR (signal-to-noise ratio) while maintaining high spatial resolution. The relationships between the datasets $X$, $Y$, and $Z$ are described as follows:
\begin{equation}
\label{eq:ch6_eq1}
X = ZS + E_s
\end{equation}
\begin{equation}
\label{eq:ch6_eq2}
Y = RZ + E_r
\end{equation}
where $S$ and $R$ are transformation matrices, $S$ is a spatial transformation matrix dictating the transformation of spatial features in $Z$ for the low dimension dataset, and $R$ is a spectral transformation matrix converting the spectral features in $Z$ to match those in the lower spectral fidelity dataset. $E_s$ and $E_r$ are residuals. The data fusion problem is to estimate $Z$, which can be done via NMF (non-negative matrix factorization):
\begin{equation}
\label{eq:ch6_eq3}
Z = WH + \epsilon
\end{equation}
where $W$ denotes spectral components, and $H$ represents corresponding abundance maps; $\epsilon$ is the residual (error) that is assumed to be zero if $W$ and $H$ are accurately obtained. 

The $X$ and $Y$ can be approximated similarly as follows:
\begin{equation}
\label{eq:ch6_eq4}
X \approx WH_h
\end{equation}
\begin{equation}
\label{eq:ch6_eq5}
Y \approx W_{m}H
\end{equation}
where $H_h$ is the spatially reduced abundance matrix, and $W_m$ is the spectrally deteriorated component matrix. 

We first unmix the LR-HS dataset to retrieve an estimate for $W$ and $H_h$; then we initialize the decomposition of the HR-LS dataset using $W$ and upsampled $H_h$ to obtain $H$. The two matrices $W$ and $H$ can be multiplied to obtain the fused dataset, $Z$. To implement this, we modified the standard version of the \href{https://scikit-learn.org/stable/modules/generated/sklearn.decomposition.NMF.html}{NMF algorithm} in the \href{https://scikit-learn.org/stable/}{scikit-learn library}, by incorporating a ``sum-to-one'' constraint on the abundance matrix. This adaptation is specifically designed for use with STEM-EDX data.

\clearpage

\input{Fig/FigS1}
\input{Fig/FigS2}
\input{Fig/FigS3}
\input{Fig/FigS4-1}
\input{Fig/FigS4-2}
\input{Fig/FigS5-1}
\input{Fig/FigS5-2}
\input{Fig/FigS6}
\input{Fig/FigS7}

\FloatBarrier

\end{document}

%% file: Fig/Fig1.tex
\begin{figure}[!htbp]
  \centering
  \includegraphics[height=14cm]{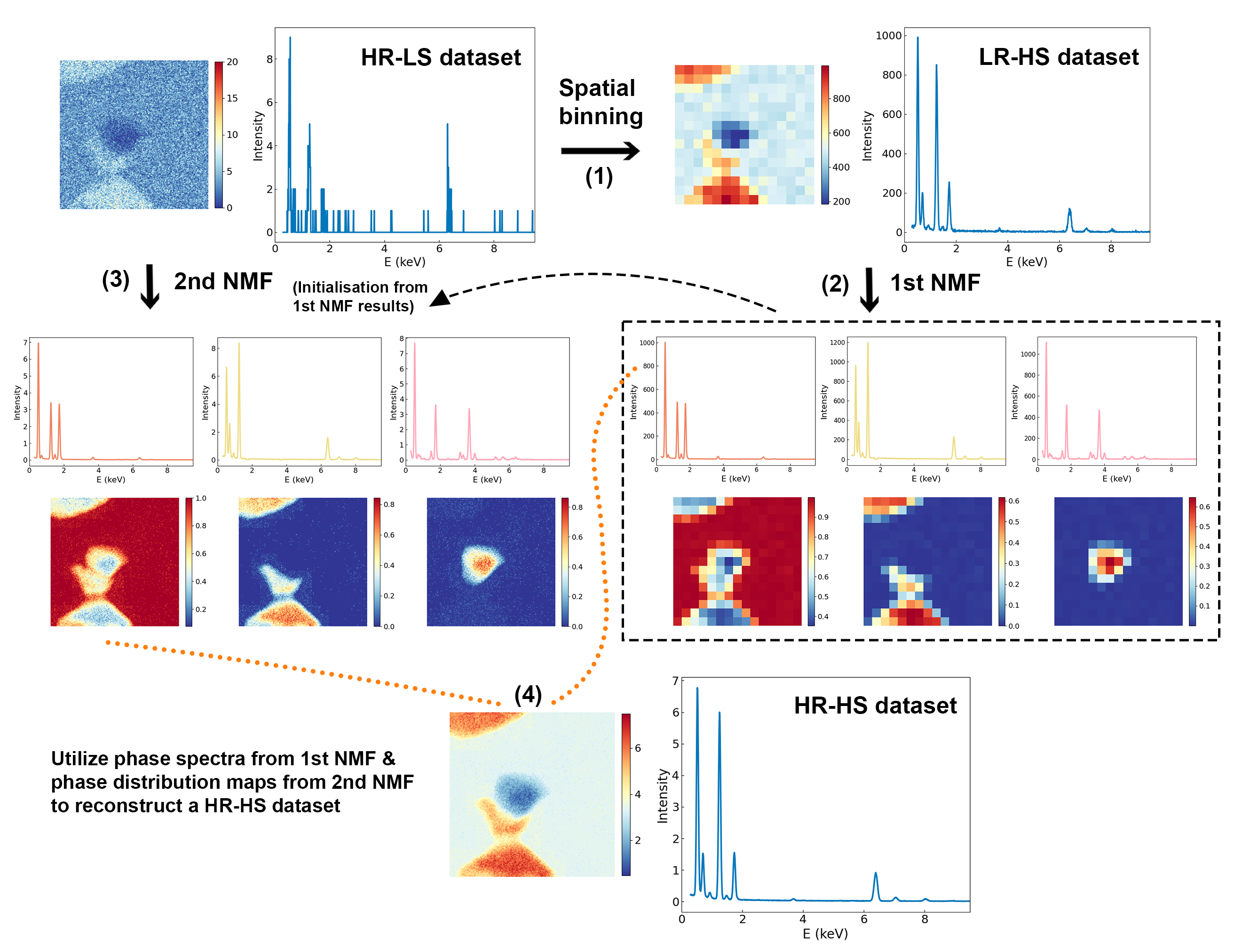}
   \caption{Schematic illustration of PSNMF methodology. From an initial STEM-EDX dataset (HR-LS), binning using spatial factor $b$ is applied to create an LR-HS dataset with improved SNR of the pixel spectra. Performing a first NMF on the LR-HS dataset generates spatial and spectral components of which the spectral components are high-quality. These are used to initialise a second NMF decomposition of the HR-LS dataset, delivering low noise, high resolution spatial components (maps). A final, high quality reconstructed dataset is made by combining the spectra of the first NMF with the spatial distribution maps of the second NMF.}
   \label{fig:PSNMF}
\end{figure}

%% file: Fig/Fig2.tex
\begin{figure}[!htbp]
  \centering
  \includegraphics[height=14cm]{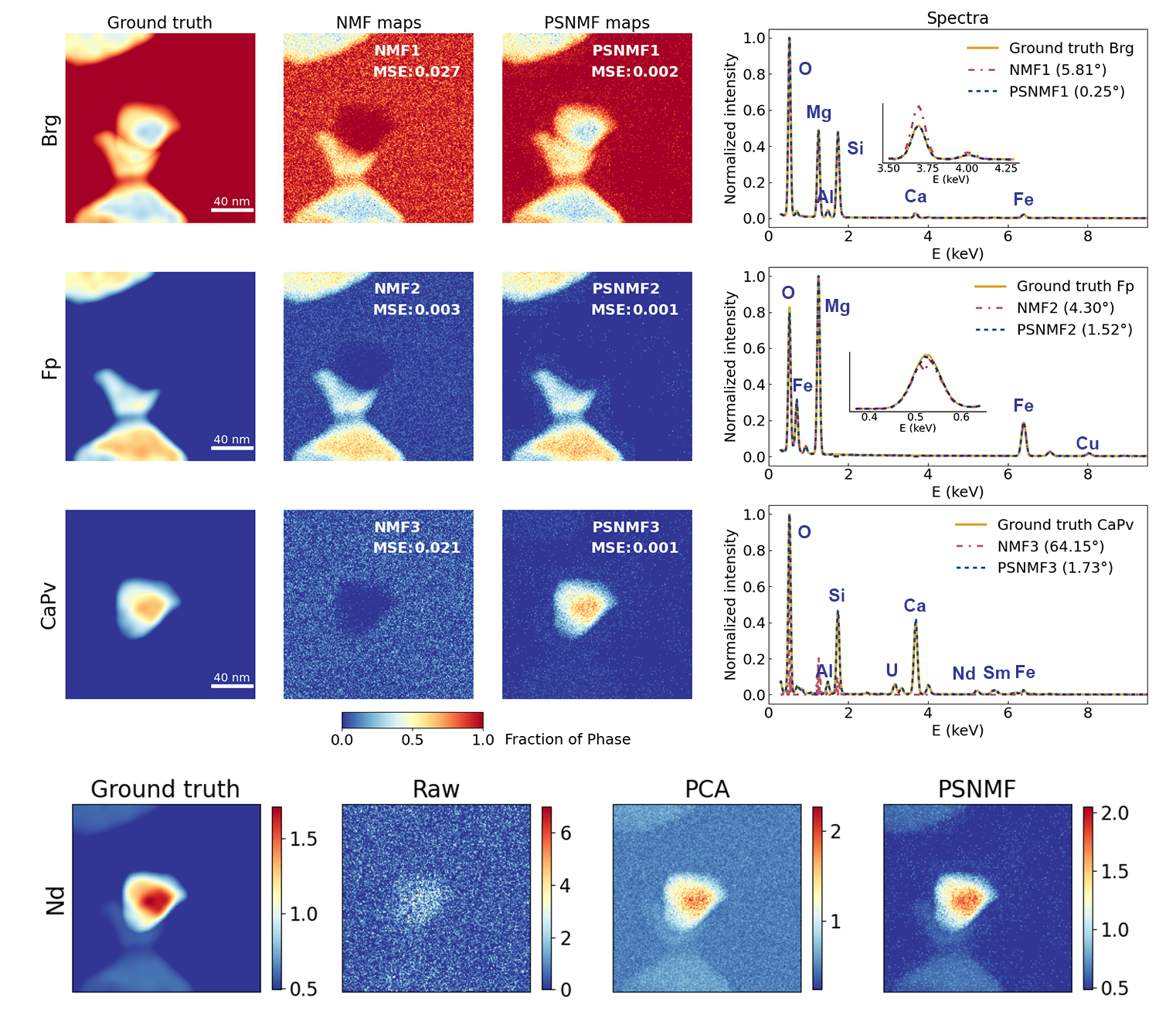}
   \caption{Application of PSNMF to a simulated dataset having medium SNR (average of 147 X-ray counts per pixel), respectively compared to standard NMF for phase identification and PCA for dataset denoising. The PSNMF uses $b = 12$. On the upper left, abundance maps of components from NMF and PSNMF are compared to ground truth phase maps, with phase spectral comparisons shown on the right. Note that the spectral legends and spatial loading maps display their respective spectral angles and MSE values. The two insets display magnified views of the Ca peaks in the Brg phase and the O peak in the Fp phase. On the bottom row, raw, PCA-denoised and PSNMF-denoised Nd maps are compared to the ground truth map.}
   \label{fig:C147}
\end{figure}

%% file: Fig/Fig3.tex
\begin{figure}[!htbp]
  \centering
  \includegraphics[height=14cm]{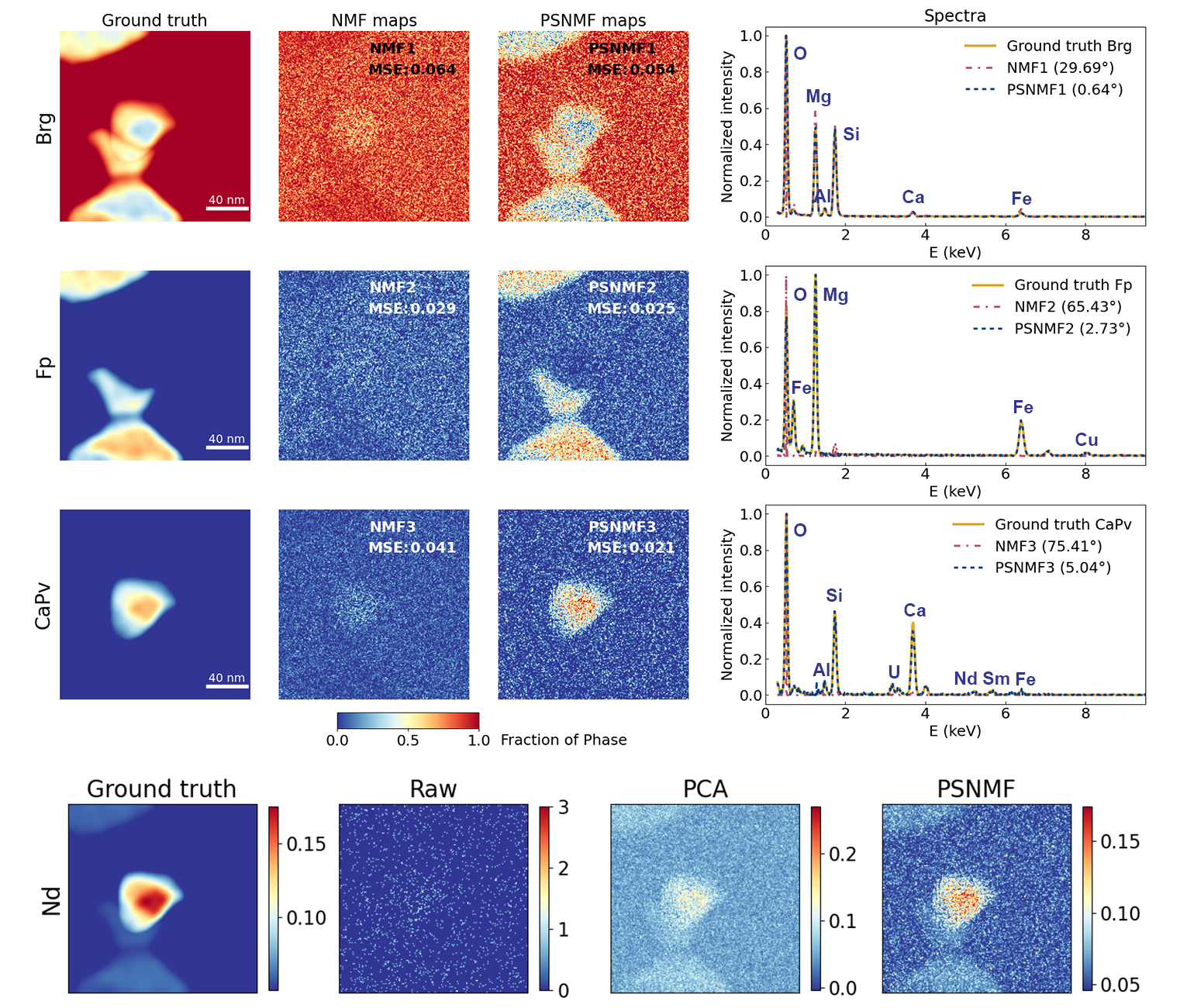}
   \caption{PSNMF applied to a simulated dataset having a very low SNR (average of 18 X-ray counts per pixel), compared to standard NMF for phase identification and to PCA for dataset denoising. The PSNMF uses $b = 15$. Abundance maps and spectra from NMF and PSNMF are compared to ground truths. On the bottom, raw, PCA-denoised and PSNMF-denoised Nd maps are compared to the ground truth map.}
   \label{fig:C15}
\end{figure}

%% file: Fig/Fig4.tex
\begin{figure}[!htbp]
  \centering
  \includegraphics[height=10cm]{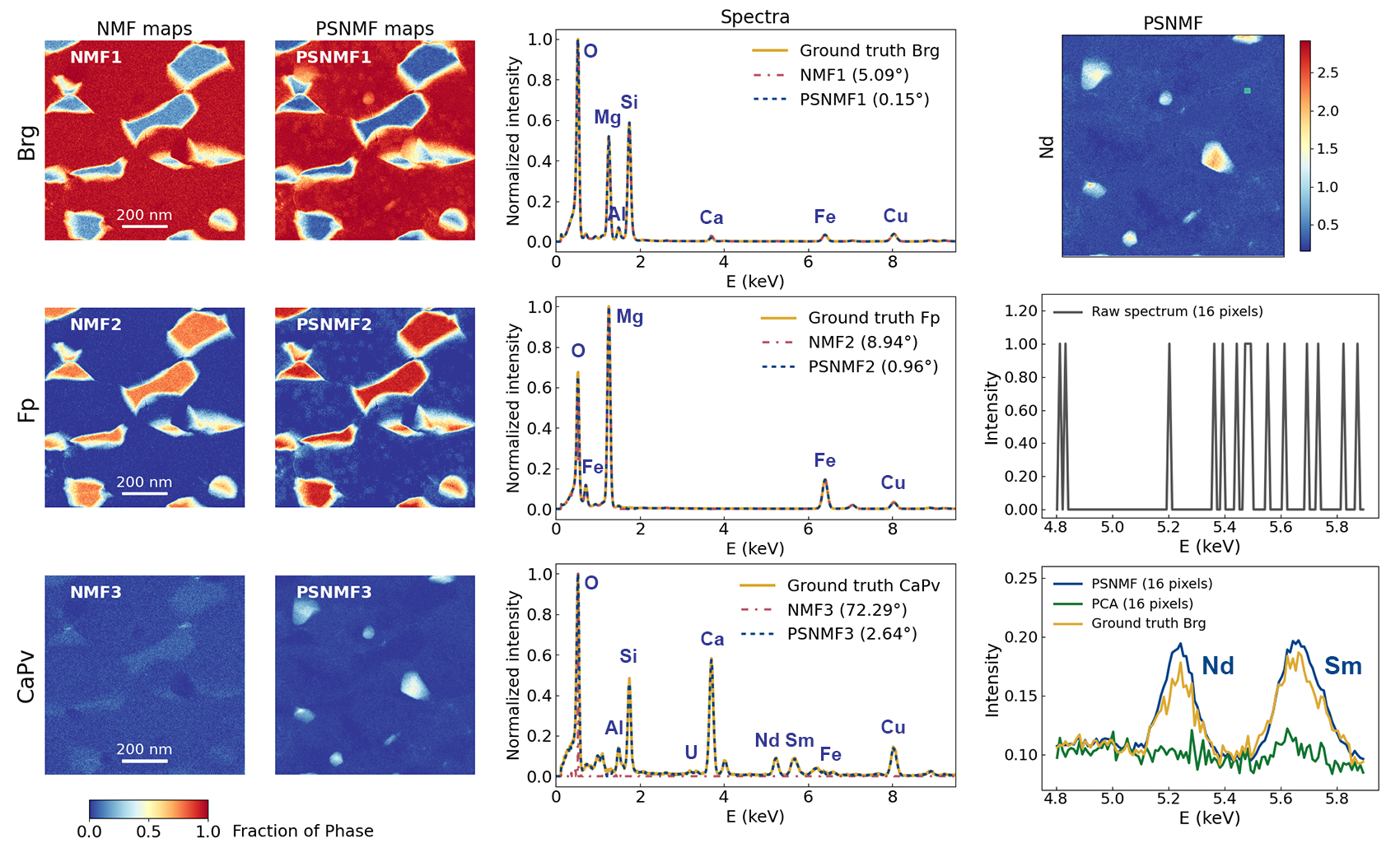}
   \caption{PSNMF applied to an experimental $512\times 512$ pixel dataset (having an average of 120 X-ray counts per pixel), acquired from a mineralogical sample. The PSNMF uses $b = 8$. The phase maps generated by PSNMF are compared to those determined using standard NMF. The phase spectra (middle column) are additionally compared to ground truth phase spectra calculated using the methodology described in Chen et al.~\cite{chen2024unmixing}. In the right column, we study the denoising effect of the PSNMF-reconstructed dataset, with the Nd map (top), and then EDX spectra integrated over the same $4 \times 4$ pixel Brg region (shown in green on the Nd map) from the raw dataset (middle), and PCA- and PSNMF-reconstructed datasets (bottom). An appropriately-scaled Brg ground truth spectrum is overlaid on the PCA/PSNMF spectra.}
   \label{fig:C120}
\end{figure}

%% file: Fig/Fig5.tex
\begin{figure}[!htbp]
  \centering
  \includegraphics[height=10cm]{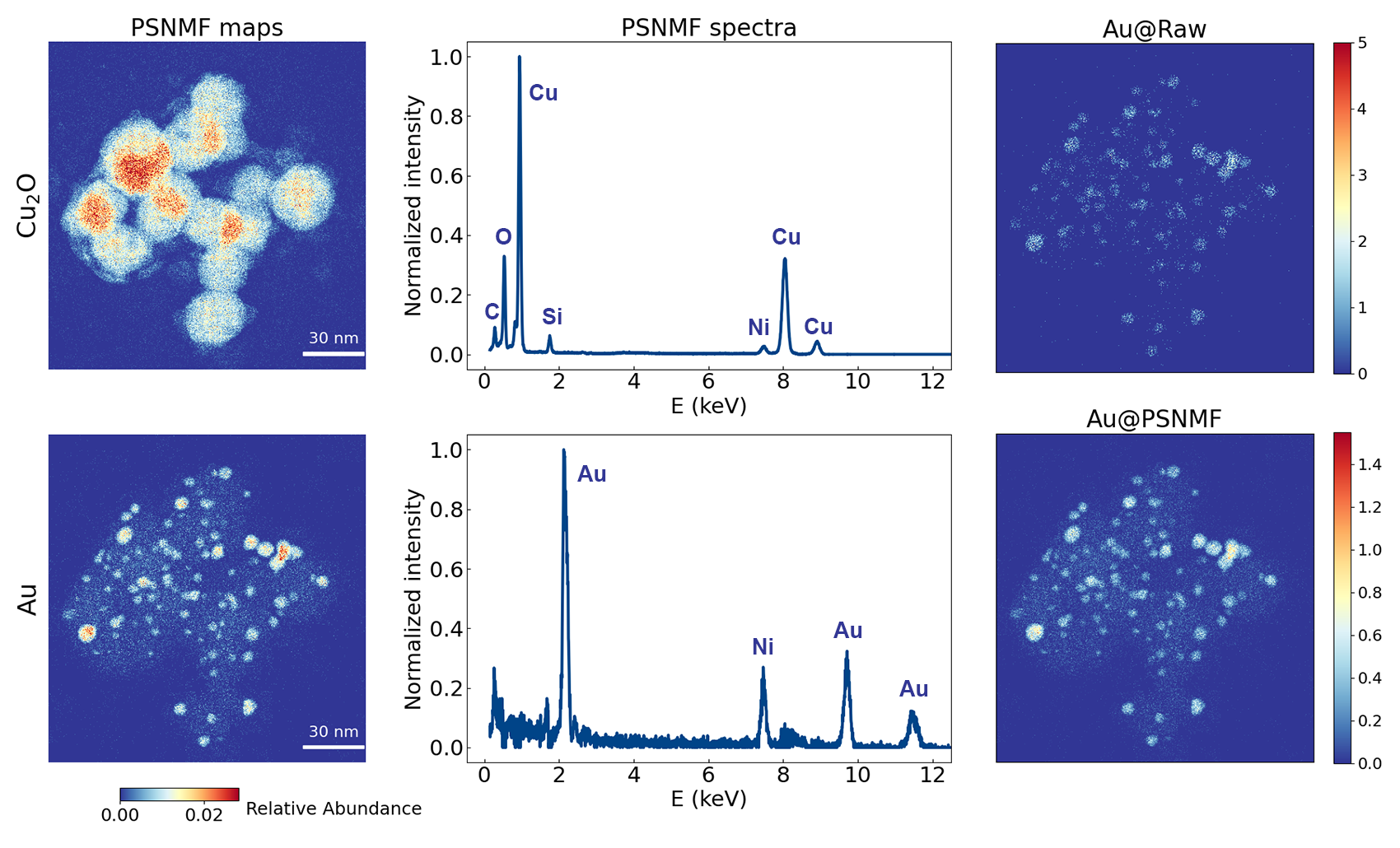}
   \caption{PSNMF applied to an experimental $512\times 512$ pixel dataset that was acquired from Au-Cu$_2$O nanoparticles supported on a carbon film. The PSNMF uses $b = 16$. Two relevant phase maps and spectra generated by PSNMF are presented. In the right column, we examine the denoising effect of the PSNMF-reconstructed dataset, as illustrated by the elemental Au map, in comparison to the raw dataset.}
   \label{fig:NPs}
\end{figure}

%% file: Fig/FigS1.tex
\begin{figure}[!htbp]
  \centering
  \includegraphics[height=6.8cm]{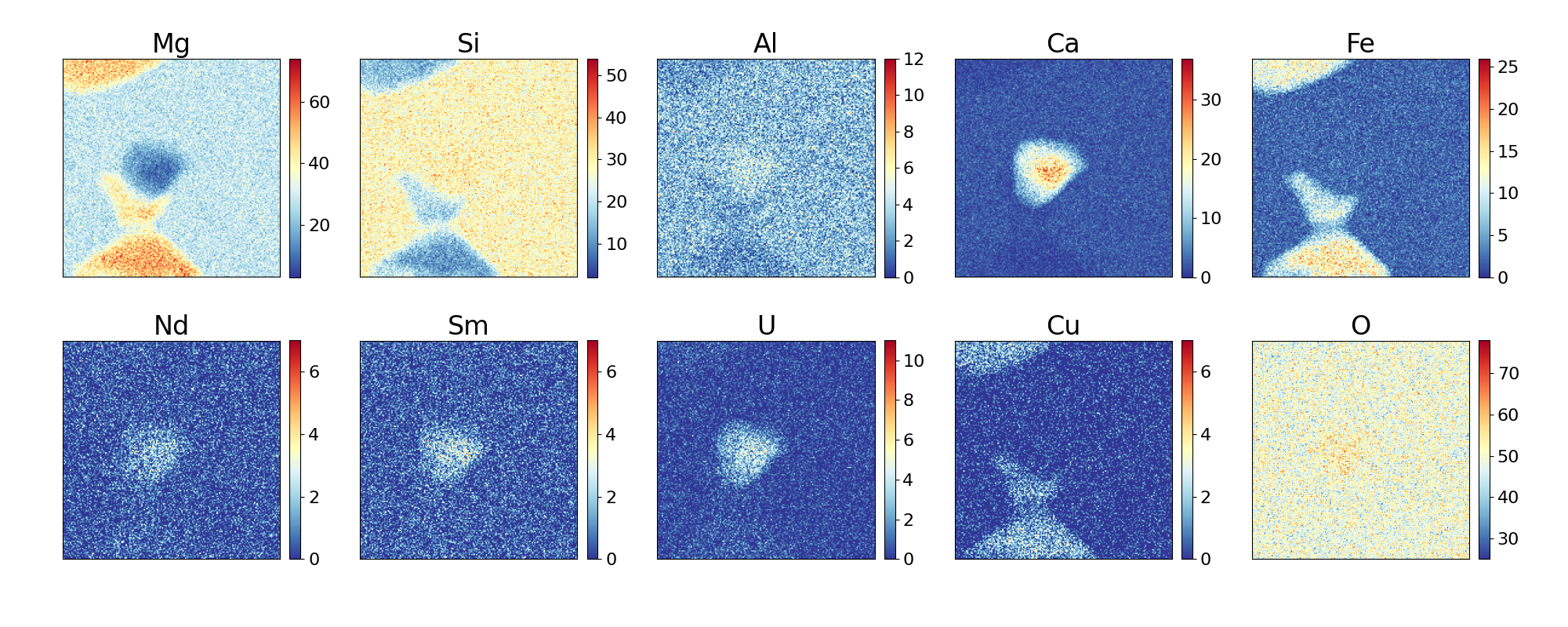}
   \caption{Raw elemental maps from the synthetic dataset with medium SNR.}
   \label{fig:c147_ele_map}
\end{figure}

%% file: Fig/FigS2.tex
\begin{figure}[!htbp]
  \centering
  \includegraphics[height=6.8cm]{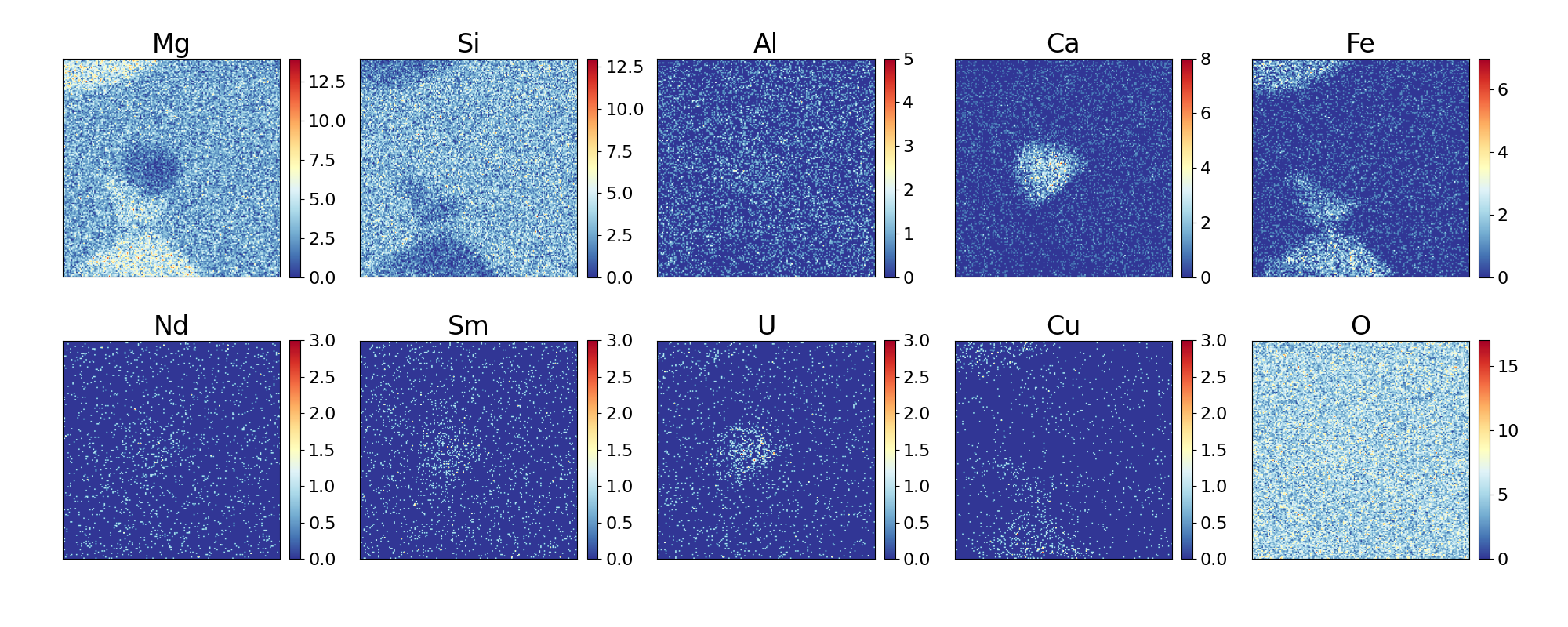}
   \caption{Raw elemental maps from the synthetic dataset with low SNR.}
   \label{fig:c15_ele_map}
\end{figure}

%% file: Fig/FigS3.tex
\begin{figure}
    \centering
    \begin{subfigure}{0.7\textwidth}{\centering\includegraphics[width=\linewidth]{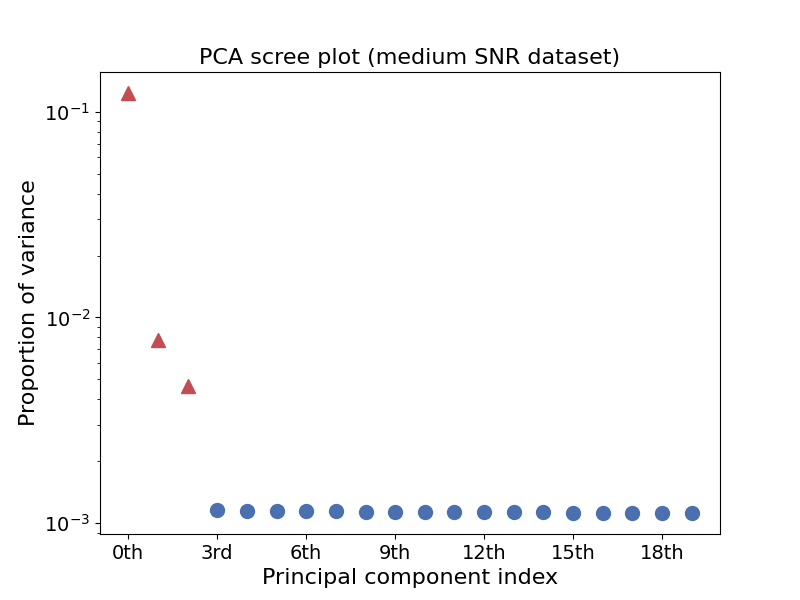}}
    \end{subfigure}
    \begin{subfigure}{0.7\textwidth}{\centering\includegraphics[width=\linewidth]{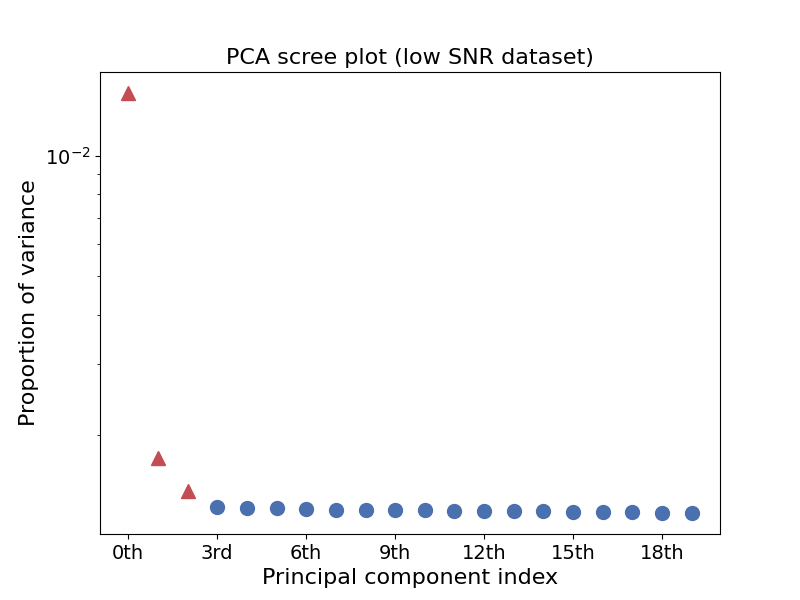}}
    \end{subfigure}
    \caption{The scree plots of PCA decomposition of the medium SNR and low SNR dataset, respectively.}
    \label{appx:c147_c15_screeplot}
\end{figure}

%% file: Fig/FigS4-1.tex
\begin{figure}[!htbp]
  \centering
  \includegraphics[height=16.0cm]{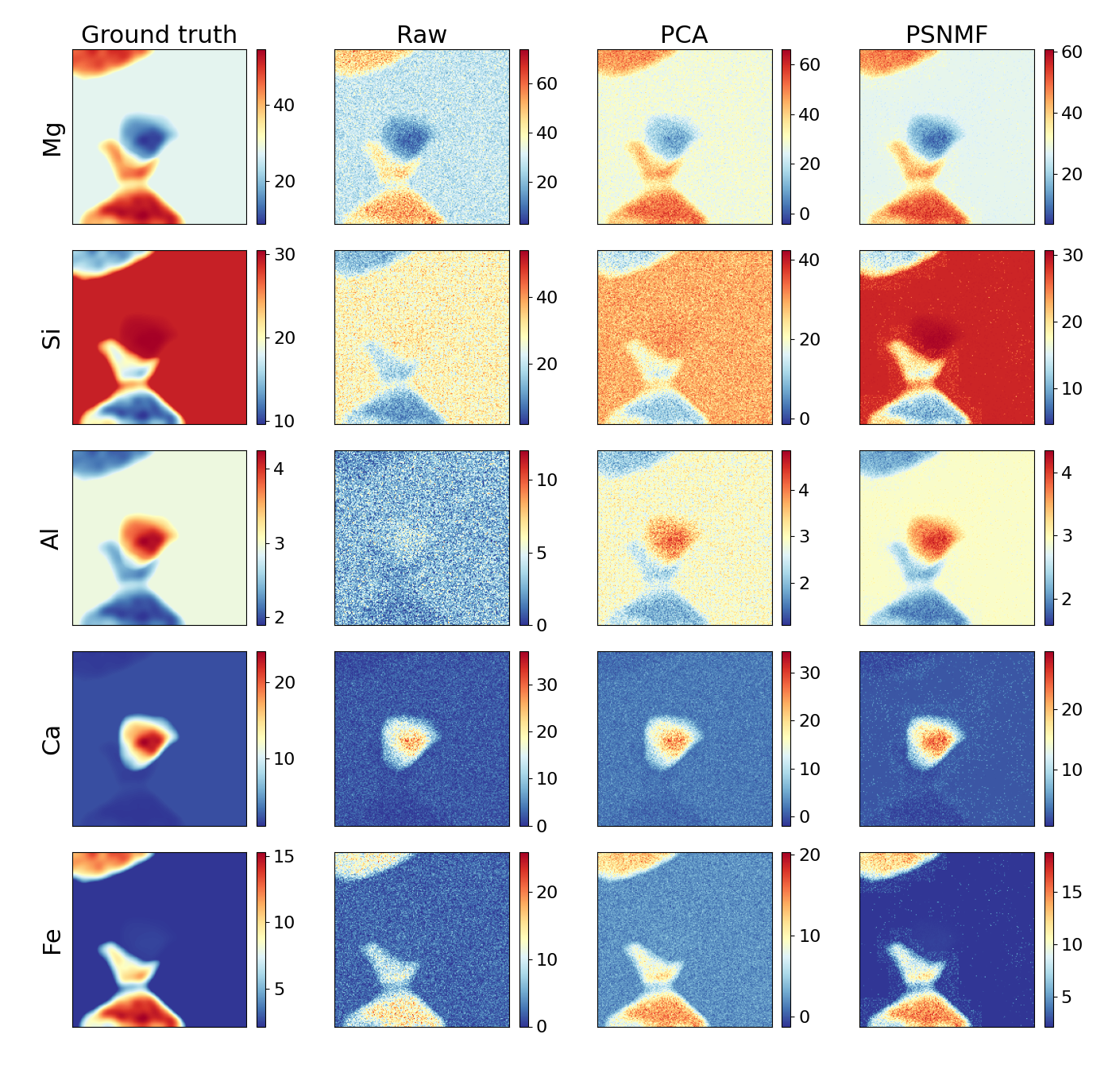}
   \caption{A full comparison of ground truth, raw, and PCA- and PSNMF-denoised elemental maps for medium SNR dataset (Part 1).}
   \label{fig:c147_comp_ele_map_p1}
\end{figure}

%% file: Fig/FigS4-2.tex
\begin{figure}[!htbp]
  \centering
  \includegraphics[height=16.0cm]{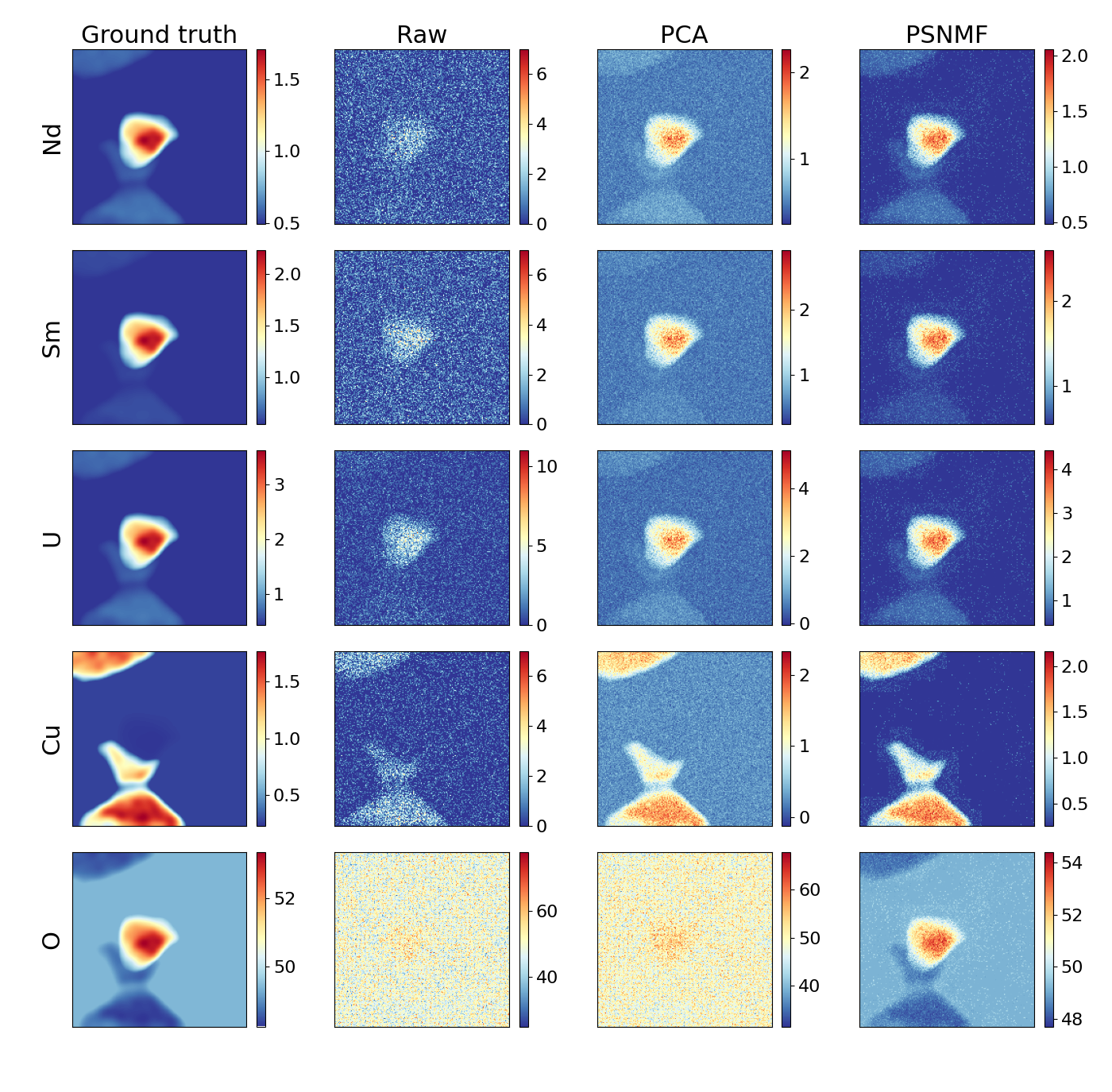}
   \caption{A full comparison of ground truth, raw, and PCA- and PSNMF-denoised elemental maps for medium SNR dataset (Part 2).}
   \label{fig:c147_comp_ele_map_p2}
\end{figure}

%% file: Fig/FigS5-1.tex
\begin{figure}[!htbp]
  \centering
  \includegraphics[height=16.0cm]{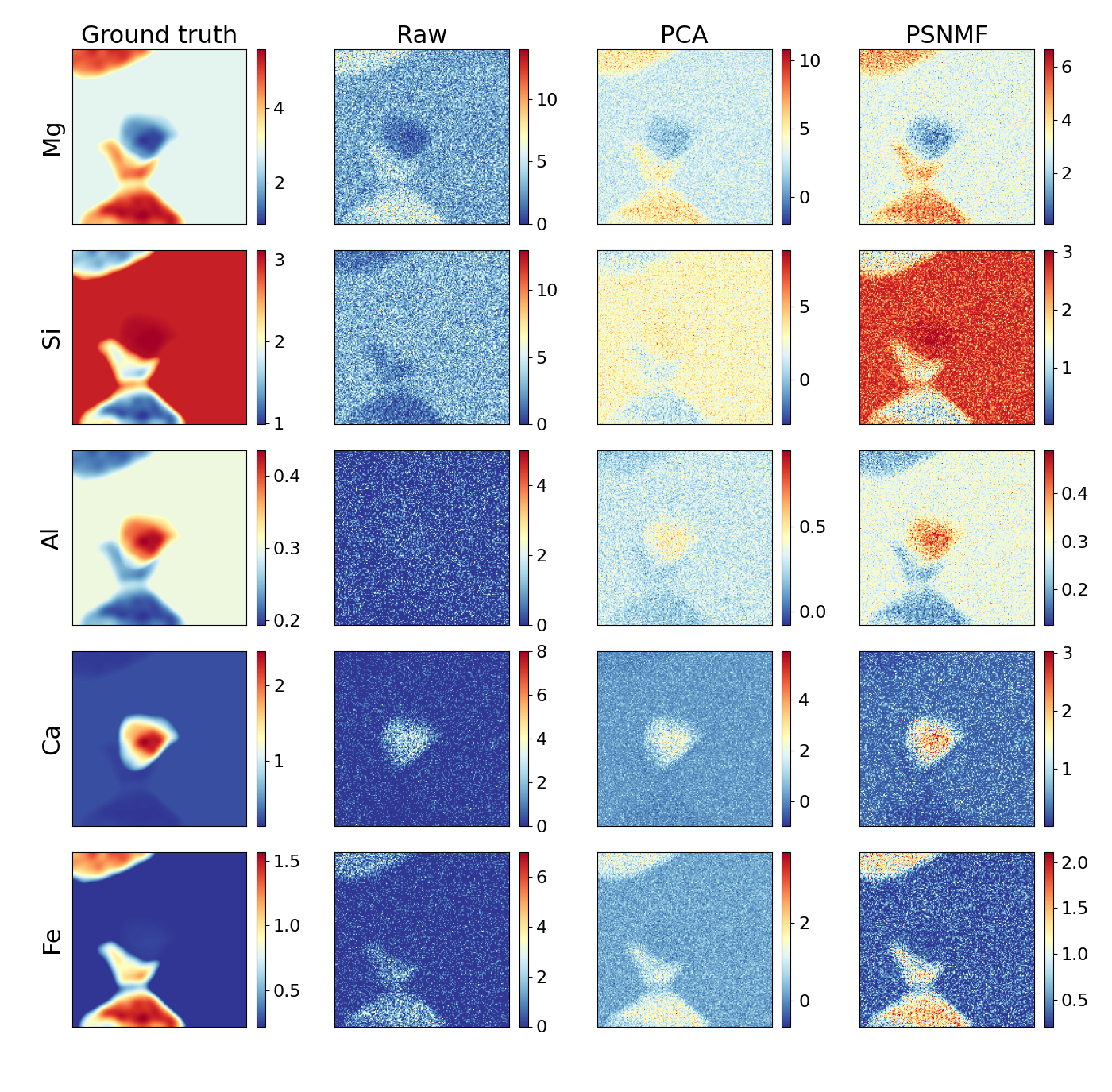}
   \caption{A full comparison of ground truth, raw, and PCA- and PSNMF-denoised elemental maps for low SNR dataset (Part 1).}
   \label{fig:c15_comp_ele_map_p1}
\end{figure}

%% file: Fig/FigS5-2.tex
\begin{figure}[!htbp]
  \centering
  \includegraphics[height=16.0cm]{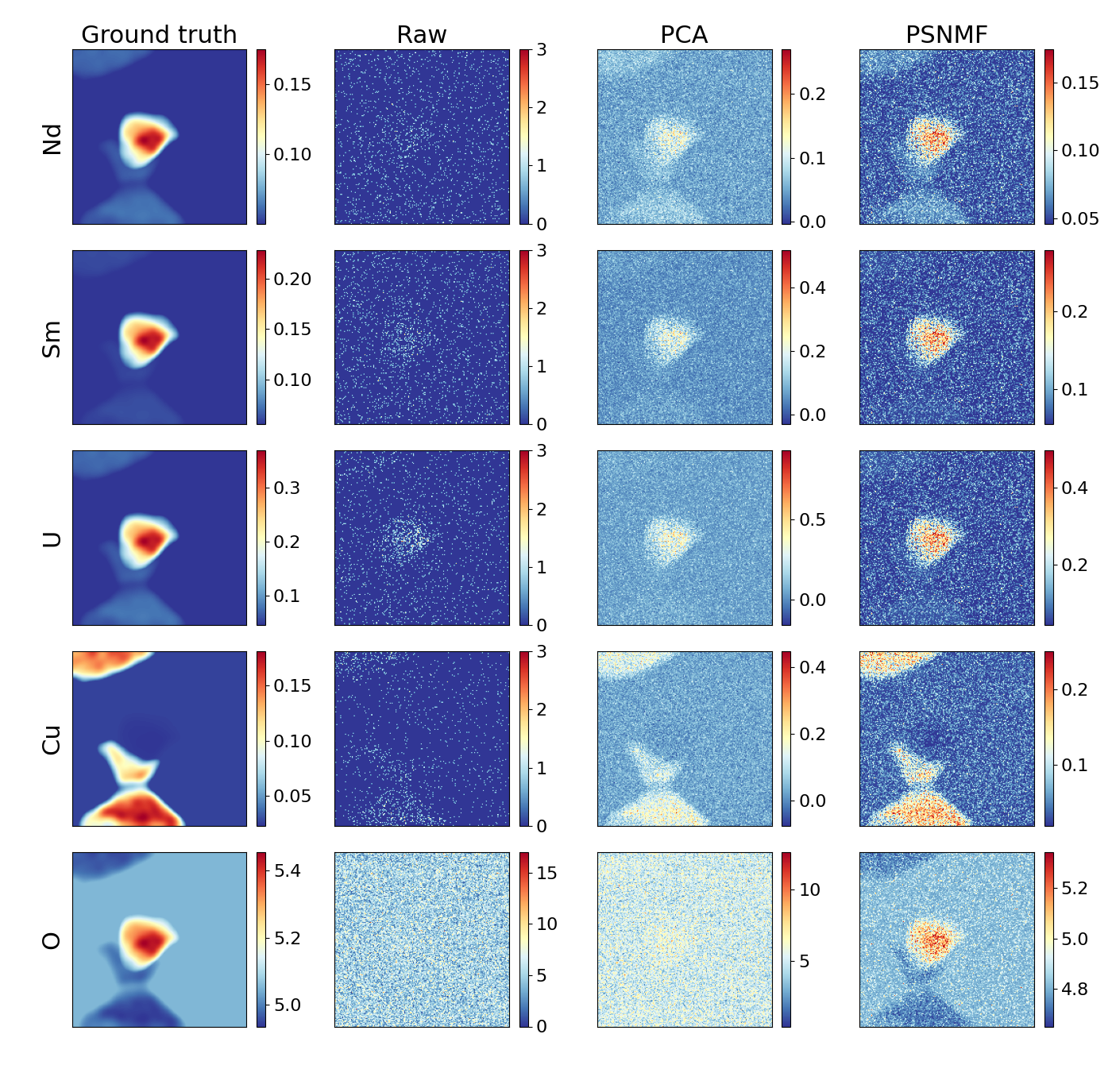}
   \caption{A full comparison of ground truth, raw, and PCA- and PSNMF-denoised elemental maps for low SNR dataset (Part 2).}
   \label{fig:c15_comp_ele_map_p2}
\end{figure}

%% file: Fig/FigS6.tex
\begin{figure}[!htbp]
  \centering
  \includegraphics[height=14cm]{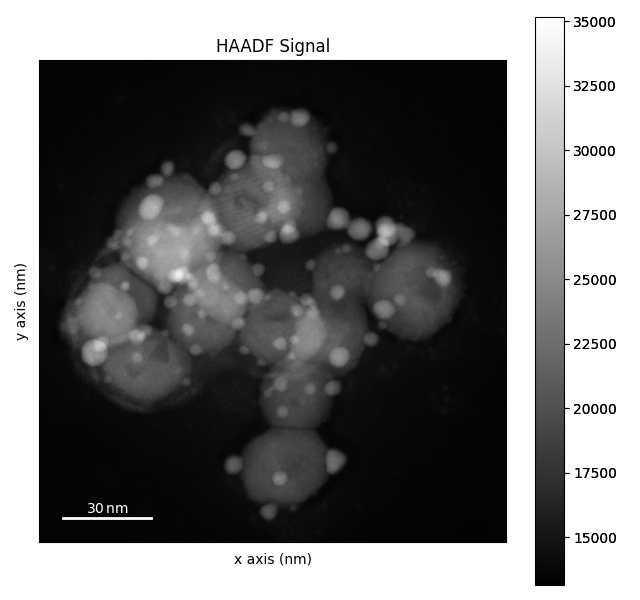}
   \caption{A high angle annular dark-field ($Z$-contrast) STEM image of the Au-Cu$_2$O nanoparticles supported on carbon.}
   \label{fig:nps_haadf}
\end{figure}

%% file: Fig/FigS7.tex
\begin{figure}[!htbp]
  \centering
  \includegraphics[height=18cm]{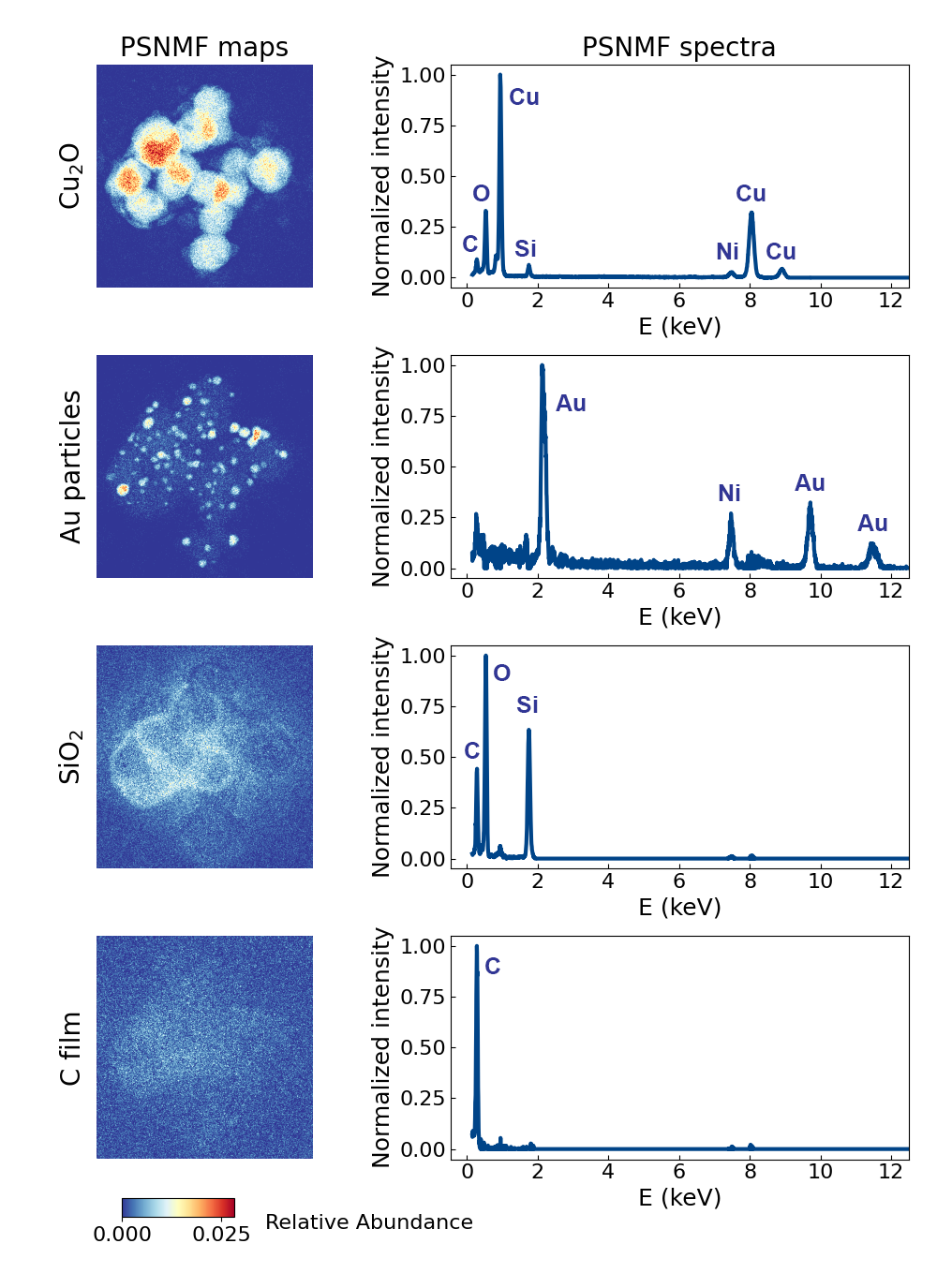}
   \caption{PSNMF decomposition results on analyzed nanoparticle sample.}
   \label{fig:nps_psnmf_decomp}
\end{figure}